\documentclass[traditabstract]{aa} 
\usepackage{graphicx}
\usepackage{txfonts}
\usepackage{url}
\usepackage[round]{natbib}
\bibpunct{(}{)}{;}{a}{}{,}

\DeclareMathSymbol{\lesssim}{\mathrel}{AMSa}{"2E}
\begin{document}
\bibliographystyle{aa}
 \title{Peering at the outflow mechanisms in the transitional pulsar PSR J1023+0038: simultaneous VLT, XMM-Newton, and Swift high-time resolution observations}
    \author{M. C. Baglio,
          \inst{1, 2}    
          \,      
          F. Vincentelli,
          \inst{3, 4, 2}
          \,
          S. Campana,
          \inst{2}
          \,      
         F. Coti Zelati,
          \inst{5, 2}
         \,
          P. D'Avanzo,
          \inst{2}
           \,      
           L. Burderi,
           \inst{6}
           \,
         P. Casella,
         \inst{4}
           \,
         A. Papitto,
         \inst{4}
         \,
         D. M. Russell
         \inst{1}
         \,
           }
		  
   \institute{New York University of Abu Dhabi, P.O. Box 129188, Abu Dhabi, UAE                      
              \email{mcb19@nyu.edu}
         \
             \and
             INAF, Osservatorio Astronomico di Brera, Via E. Bianchi 46, I-23807 Merate (LC), Italy
           \and
           Department of Physics \& Astronomy, University of Southampton, Highfield, Southampton SO17 1BJ, UK
           \and
           INAF, Osservatorio Astronomico di Roma, Via Frascati 33, I-00078 Monteporzio Catone (Roma), Italy
           \and
           Institute of Space Sciences (ICE, CSIC-IEEC), Campus UAB, Carrer Can Magrans s/n, E-08193 Barcelona, Spain
           \and
           Universit\`{a} degli Studi di Cagliari, Dipartimento di Fisica, SP Monserrato-Sestu, KM 0.7, I-09042 Monserrato (CA), Italy
                           }
   \date{ }

   \abstract{ 
   We report on a near infrared, optical and X-ray simultaneous campaign performed in 2017 with the \textit{XMM-Newton} and \textit{Swift} satellites and the HAWK-I instrument mounted on the VLT on the transitional millisecond pulsar PSR J1023+0038. Near infrared observations were performed in fast-photometric mode (0.5s exposure time) in order to detect any fast variation of the flux and correlate them with the optical and X-ray light curves. The optical light curve shows the typical sinusoidal modulation at the system orbital period (4.75hr). No significant flaring or flickering is found in the optical, neither signs of transitions between active and passive states. On the contrary, the near infrared light curve displays a bimodal behaviour, showing strong flares in the first part of the curve, and an almost flat trend in the rest. The X-ray light curves instead show a few \textit{low}/\textit{high} mode transitions, but no flaring activity is detected. Interestingly, one of the \textit{low}/\textit{high} mode transition is found to happen at the same time as the emission of an infrared flare. This can be interpreted in terms of the emission of an outflow or a jet: the infrared flare could be due to the evolving spectrum of the jet, which possesses a break frequency that moves from higher (near infrared) to lower (radio) frequencies after the launching, that has to happen at the \textit{low}/\textit{high} mode transition. We also present the cross correlation function between the optical and near infrared curves. Due to the bimodality of the near infrared curve, we divided it in two parts (flaring and quiet). While the cross correlation function of the quiet part is found to be flat, the one referring to the flaring part shows a narrow peak at $\sim$10s, which indicates a delay of the near infrared emission with respect to the optical. This lag can be interpreted as reprocessing of the optical emission at the light cylinder radius with a stream of matter spiraling around the system due to a phase of radio-ejection. This strongly supports a different origin of the infrared flares observed for PSR J1023+0038 with respect to the optical and X-ray flaring activity reported in other works on the same source.

 }

\authorrunning{M. C. Baglio et al.} 
\titlerunning{Peering at the outflow mechanisms in the transitional pulsar PSR J1023+0038}
\maketitle

\section{Introduction}\label{intro}
Millisecond pulsars (MSPs) are neutron stars (NSs) emitting pulsed radiation (typically in the radio band, but also in the X- and $\gamma$-rays) with a period within the 1-10 milliseconds range. According to the \textit{recycling scenario}, MSPs would be NSs originally hosted in a binary system (typically a low-mass X-ray binary - LMXB) that have been re-accelerated through accretion. In particular, the angular momentum transfer due to accretion could increase the rotational velocity of the NS around its axis up to hundred times per second or more, as observed in MSPs (\citealt{Alpar82}; \citealt{Radhakrishnan82}; \citealt{Srinivasan2010}). 

A confirmation to the recycling scenario of MSPs came first with the discovery of ms X-ray pulsations from the NS LMXB SAX J1808.4-3658 \citep{Wijnands}, and more recently with the so-called \textit{missing link} pulsar, PSR J1023+038 (hereafter J1023), which was the first source that was discovered to transit between a radio pulsar state and a LMXB state \citep{Archibald2009}, giving birth to the class of the transitional millisecond pulsars (tMSPs). The first system showing instead a full outburst cycle, going through a bright X-ray phase ($\sim 10^{36}\, \rm erg\,\rm s^{-1}$) down to quiescence was IGR J18245-2452 \citep{Papitto2013}. Nearly two weeks after settling to quiescence, a radio pulsar started shining, thus confirming the MSP nature of the compact object.

J1023 was first detected by \citet{Bond2002} in the radio band, and, at a later time, showed in the optical clear signatures for the presence of an accretion disc around the compact object \citep{Szkody2003}. Later on, \citet{Thorstensen2005} identified J1023 as a  NS-LMXB. Observations carried out in 2003-2005 however revealed no presence of the accretion disc, the spectrum being consistent with that of a G-type star, but instead signatures of strong irradiation were detected \citep{Thorstensen2005}. In 2007, the compact object of the system was identified as a 1.69 ms period radio pulsar \citep{Archibald2013}. Finally, J1023 underwent a state change from the MSP state to a LMXB state (where it has remained until now in 2019) in June 2013, and the radio pulsar signal consequently switched off \citep{Stappers2014}. 
A large number of multi-wavelength studies have been performed on this source since its last state transition (see e.g. \citealt{Takata2014}; \citealt{Tendulkar2014}; \citealt{CotiZelati2014}; \citealt{Deller2015}; \citealt{Bogdanov2015}; \citealt{Shahbaz2015}; \citealt{Baglio2016b}; \citealt{Campana2016}; \citealt{Jaodand2016}; \citealt{Ambrosino2017}; \citealt{Papitto2018}; \citealt{CotiZelati2018}; \citealt{Bogdanov2018}; \citealt{Shahbaz2018}).

While in the LMXB state, a very peculiar behaviour in the X-rays is revealed  \citep{Bogdanov2015}: J1023 exhibits frequent mode switchings between three different X-ray states. For about $\sim 70\%$ of the time, a \textit{high} luminosity mode ($L_{X} \sim 7\times 10^{33}$ \, erg s$^{-1}$, 0.3-79 keV; \citealt{Tendulkar2014}) occurs, which is characterized by the observation of coherent X-ray pulsations \citep{Archibald2015}, with an rms pulsed fraction of $\sim 8\%$. Another $20\%$ of the time, the system is in a \textit{low} luminosity mode, characterized by $L_{X} \sim 5\times 10^{32} \, \rm erg \, s^{-1}$ and by the absence of X-ray pulsations. Finally, for $\sim$ 2$\%$ of the time, a \textit{flaring} mode occurs, during which a sudden increase of the X-ray luminosity up to $3 \times 10^{34} \rm erg \, s^{-1}$ is reported (\citealt{Bogdanov2015}). Luminosities are calculated at a distance of 1.37 kpc, as derived from radio parallax \citep{Deller2015}. Transitions between the \textit{high} and the \textit{low} luminosity modes happen on a time-scale of $\sim 10$s, and are visualized in the X-ray light curve as rectangular-shaped. Similarly, at optical frequencies frequent transitions between two different modes (named \textit{active} and \textit{passive} mode) were first observed by \citet{Shahbaz2015} as rectangular dips in the light curves, with ingress and egress times of $\sim 20$s. However, the existence of a mode-switching at optical frequencies is, at the moment, strongly debated.
Bright emission in the radio with a flat to slightly inverted spectrum was detected by \citet{Deller2015} and was interpreted as due to the emission of a compact, self-absorbed jet. Later, \citet{Bogdanov2018} observed the emission of radio flares in correspondence with the X-ray \textit{low} mode, while during the \textit{high} mode a steady radio emission was detected. This strong anti-correlation between the radio and the X-ray light curve was again interpreted as due to synchrotron emission from jets or outflows.
Optical, phase-resolved linear polarization measurements were also performed (\citealt{Baglio2016b}; \citealt{Hakala2018}), revealing a degree of polarization of a few per cent which is better interpreted with being due to Thomson scattering from the accretion disc than to synchrotron emission from a jet.
Strong flickering and flaring activity was also observed in the near-infrared (\citealt{Hakala2018}; \citealt{Shahbaz2018}), and interpreted as due to reprocessing of the optical emission and to a combination of reprocessing and emission from plasmoids in the accretion flow in the near infrared. In addition, \textit{Kepler} monitoring found a much stronger flaring activity in the optical than in the X-ray band (see \citealt{Papitto2018}).

Very recently, optical pulsations at the pulsar spin period were observed during the accretion state, making J1023 the first optical millisecond pulsar ever detected (\citealt{Ambrosino2017}; see also \citealt{Zampieri2019}). This puzzling behaviour was interpreted as due to the presence of an active rotation-powered millisecond pulsar in the system.

Several models have been proposed in order to explain the unprecedented variety of features observed for this system. Some of these models involve the presence of an active pulsar which is enshrouded by the accretion disc or by a plasma of ionized matter, causing the interaction between the pulsar wind and the ionized matter and, therefore, the emission of synchrotron radiation in the X-rays (\citealt{Takata2014}; \citealt{CotiZelati2014}; \citealt{Li2014}); others see the presence of a fast spinning NS which is propelling away the matter coming from the accretion disc \citep{PapittoTorres2015}; still others consider instead the alternation between a radio pulsar and a propeller regime, which might account for the continuous switches between the low and the high X-ray modes (\citealt{Campana2016}; \citealt{CotiZelati2018}). Recently, \citet{Papitto2019} suggested that J1023 might be a rotation-powered pulsar emitting a magnetized wind which interacts with the accreting mass close to the light cylinder radius, therefore creating a shock. At this shock, electrons are accelerated, producing optical and X-ray pulsations that have a synchrotron nature. This interpretation also accounts for the lack of pulsations in the X-ray and optical in the \textit{low} mode, by hypothesizing that the shock is moved at larger distance by the pulsar wind, and of the \textit{flaring} mode, considering that the shock region might increase its size.

In this paper we show the results of a simultaneous near-infrared/optical/X-ray fast imaging campaign performed with the Very Large Telescope + \textit{XMM-Newton} and \textit{Swift} on PSR J1023+0038 on June 9-10, 2017. Fast near-infrared photometry is performed simultaneously to X-ray observations which allowed us to link the fast variability of the near-infrared light curve to variations in the X-ray (and optical) emission. The details of the observation and the data analysis are reported in Sec. \ref{obs_dataanalysis_sec}; in Sec. \ref{Sec_results} the main results of the campaign are shown; the discussion and conclusions are drawn in Sec. \ref{Sec_discussion} and \ref{Sec_conclusions}.

\section{Observation and data analysis}\label{obs_dataanalysis_sec}
\subsection{Near-infrared fast-photometry}
We collected near-infrared ($J$-band, 1.2 $\mu$m) high time resolution data with the HAWK-I instrument mounted on the Very Large Telescope (VLT), UT-4/Yepun, on June 10th, 2017. HAWK-I is a near-infrared (NIR) wide-field imager (0.97 to 2.31 $\mu$m) made by four HAWAII 2RG 2048x2048 pixels detectors displayed in a $2\times2$ configuration \citep{Kissler2008}. Each quadrant has a field of view (FoV) of  217 arcsec$^2$ and it is separated by the nearby detector by $15''$ gap. The total FoV is therefore $7.5' \times 7.5 '$ In order to reach sub-second time resolutions, observations were performed in \textit{Fast Phot} mode: such mode limits the area on which the detector is read to one stripe in each quadrant. In particular, we set the instrument to read 16 contiguous windows of 128 x 256 pixels. HAWK-I was pointed to (RA, DEC) 10:23:57.0, +00:37:34.3 with a rotation angle of -17.8$^\circ$ in order to place the source and a reference star (with known Vega $J$-band magnitude of $13.51\pm0.03$) in the upper-right quadrant (Q3). During the observation, a series of 100 frames are stacked together and form a ``data-cube'': each frame has a time resolution of 0.5 s. A gap of $\approx$3 seconds is present between each cube. Photometric data were extracted using the ULTRACAM data reduction software tools \citep{Dhillon2007}. For the light curve extraction, an aperture radius of $1.3''$ was used, while for the background we used an annulus going from $1.9''$ to $2.7''$ around the source. Parameters for the extraction were derived from the reference star (RA, DEC = 10:23:43.31, +00:38:19.1) and position. To account for seeing effects, the ratio between the source and the reference star count rate was used. The time of each frame was then placed into the Barycentric Dynamical Time system, and the photon arrival times were referred to the Solar System barycenter.

\subsection{X-ray observations}
\subsubsection{XMM-Newton observations}
{\it XMM-Newton} observed J1023 on June 9, 2017. 
Due to a technical problem, no data were gathered from the pn camera. The MOS1 camera (in small window mode) was exposed for 1.3 ks, and the MOS2 camera (in small window mode and overlapping with MOS1 data) for 10 ks, out of 124 ks requested, owing to anomalies of the instruments during the observation. MOS2 data however were collected only for the first 420 s. We do not consider EPIC data in the following.

The RGSs \citep{herder01} were configured in the standard spectroscopy
mode (in the 0.35--2 keV energy range) and observed the target for $\sim24$\,ks. We processed the data of each RGS instrument using the task \texttt{rgsproc}, and extracted a combined RGS1 and RGS2, first and second order, exposure-corrected and background-subtracted light curve using \texttt{rgslccorr}. We adopted a time bin of 140\,s so as to single out the high and low X-ray modes.

\subsubsection{\textit{Swift}/XRT observations}
{\it Swift} XRT started observing J1023 on June 9, 2017 at 22:34:04. The XRT was operated in Photon Counting mode (2.5s time resolution). The observation lasted 7.0 ks, split into five consecutive orbits. Data were processed with {\tt xrtpipeline} to derive a cleaned event file. A light curve in the 0.3--10 keV energy band has been extracted from a circular region with a 35 pixels radius. The source count rate is $\sim 0.3$ counts s$^{-1}$. The background count rate amounts to $1.4\%$ of the source rate. The UVOT was operated in image mode with the $m2$ filter. These data are not considered here.

\subsection{Optical observations}
The {\it XMM-Newton} Optical Monitor (OM) observed J1023 starting on June 9th at 22:32:27 UTC in fast mode, using the $B$ filter (central effective wavelength of 450 nm). Data were taken into five subsequent exposures lasting 4387.5 s each in fast mode (0.11s time resolution).
The data set ended on June 10 2017, 05:02:13, well before the expected end.
Data were processed using {\tt omfchain}. J1023 is well detected with a background subtracted mean count rate of $\sim 7$ count s$^{-1}$.

\section{Results}\label{Sec_results}



\begin{figure*}
\centering
\includegraphics[scale=0.7]{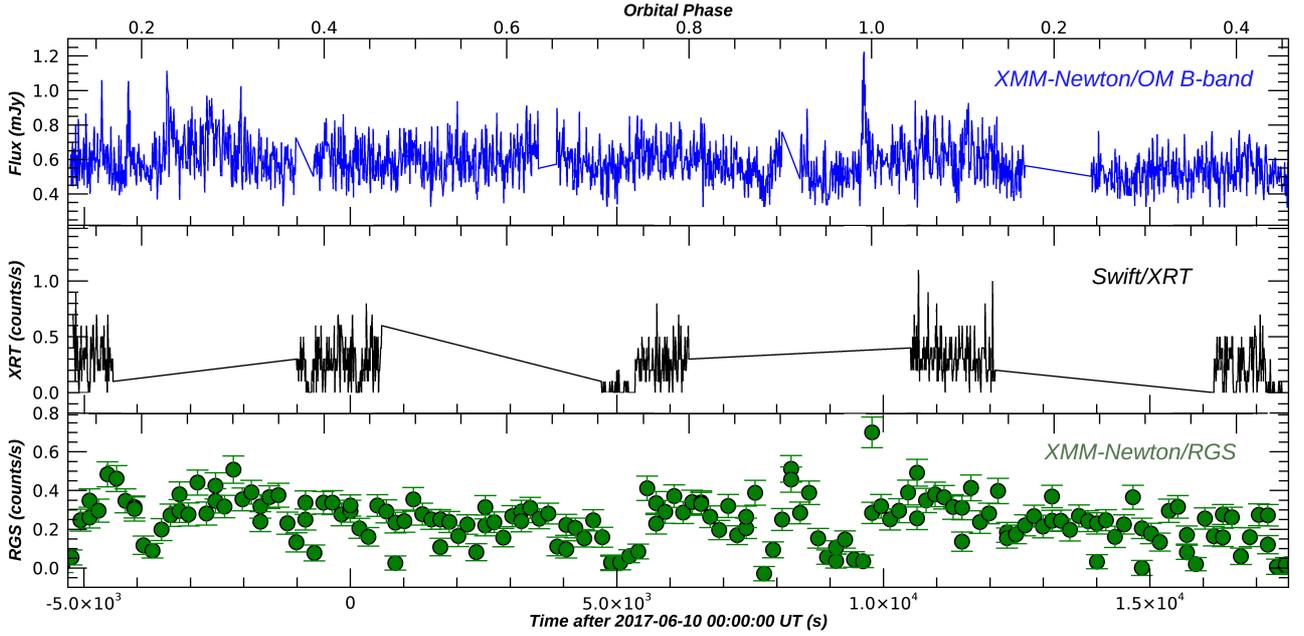}
\caption{From top to bottom, $B$-band (OM), \textit{Swift}/XRT and \textit{XMM-Newton} RGS light curves of PSR J1023 (bin time of 10s). The optical light curve has been de-trended. Errors are reported at the $68\%$ confidence level, and are shown for the RGS light curve only. For the OM and the \textit{Swift}/XRT curve, errors ($68\%$ confidence level) are of the order of 0.1 mJy and 0.2 counts/s, respectively. The bin time is 10s for the $B$-band and the \textit{Swift}/XRT curves, and 140s for the \textit{XMM-Newton} RGS curve. The top X axis reports the orbital phases of the system, which have been evaluated starting from the ephemeris of \citet{Archibald2009}, phase 0 being the inferior conjunction of the companion star.}
\label{opt_X_lc}

\end{figure*}

\begin{figure*}
\centering
\includegraphics[scale=0.7]{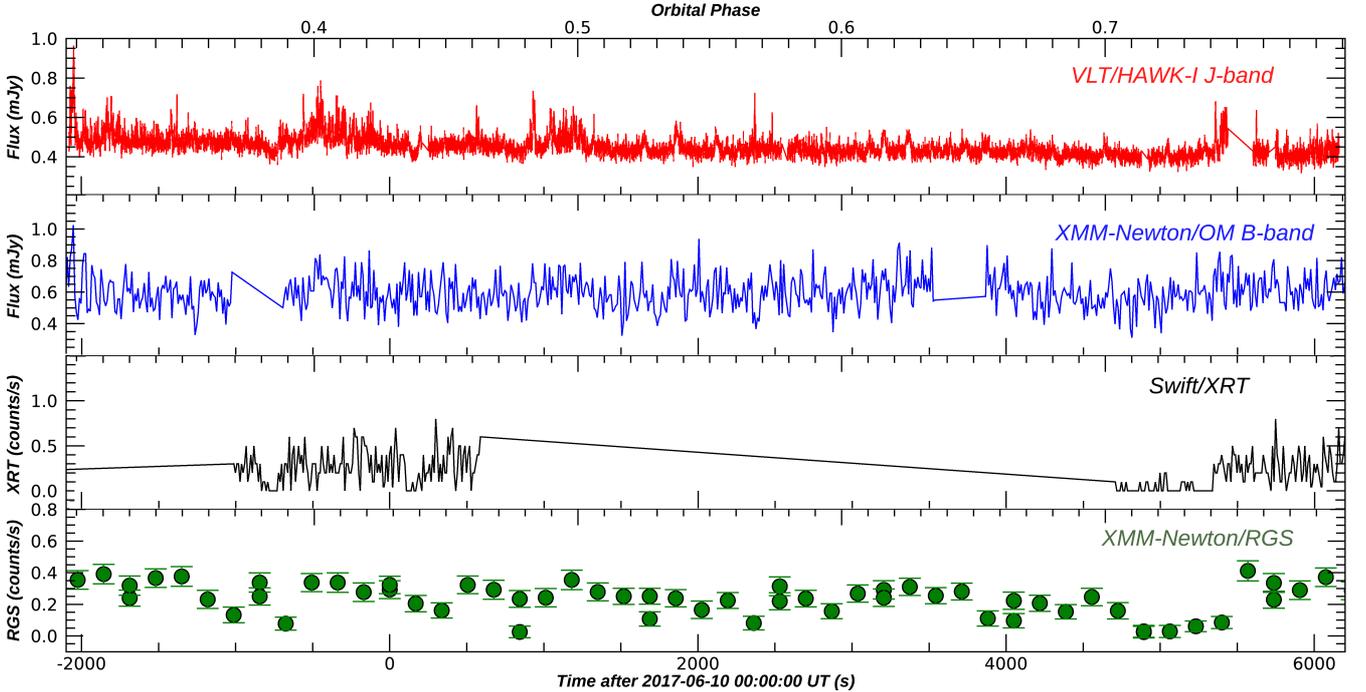}
\caption{From top to bottom, $J$-band (HAWK-I), $B$-band (OM), \textit{Swift}/XRT and \textit{XMM-Newton} RGS light curves of PSR J1023. The bin time is 0.5s for the $J$-band curve, 10s for the $B$-band and the \textit{Swift}/XRT curves, and 140s for the \textit{XMM-Newton} RGS curve. The optical light curve has been detrended. Errors are reported at the $68\%$ confidence level, and are shown for the RGS light curve only. For the HAWK-I, OM and the \textit{Swift}/XRT curves, errors ($68\%$ confidence level) are of the order of 0.02 mJy, 0.1 mJy and 0.2 counts/s, respectively. The top X axis reports the orbital phases of the system, which have been evaluated starting from the ephemeris of \citet{Archibald2009}, phase 0 being the inferior conjunction of the companion star.}
\label{NIR_opt_X_lc}

\end{figure*}

\subsection{Light curves}
In Fig. \ref{opt_X_lc} and \ref{NIR_opt_X_lc}, the simultaneous NIR (HAWK-I $J$-band), optical (XMM-Newton OM $B$-band) and X-ray (\textit{Swift}/XRT and XMM-Newton RGS) light curves of J1023 are presented. We calculated the orbital phases based on the ephemeris of \citet{Archibald2009}. Phase 0 corresponds to the inferior conjunction of the companion, i.e. when the observer sees the non-irradiated face of the donor star. The sinusoidal modulation (with a semi-amplitude of $\sim 9-10\%$) at the known 4.75 hr orbital period of the source has been detrended from the original $B$-band light curve, leaving an optical curve which still shows signs of erratic variability (Fig. \ref{opt_X_lc}, upper-panel). 
The $J$-band light curve does not clearly show any sinusoidal variability at the system orbital period (Fig. \ref{NIR_opt_X_lc}, top panel), consistent with previous works on the source at near-infrared wavelengths (see \citealt{CotiZelati2014}; \citealt{Shahbaz2018}). The expected $J$-band curve semi-amplitude, considering our $B$-band measured semi-amplitude and the values reported in \citet{CotiZelati2014} as comparison, is of $\sim 5\%$, which is of the order of the error on the single flux measurement reported in our $J$-band light curve. We therefore conclude that any sinusoidal modulation of our $J$-band light curve is likely hidden in the intrinsic statistical uncertainty of our data. Instead, a frequent flaring activity is observed, superimposed on an overall constant behaviour of the curve. This is reminiscent of what is observed in the NIR by \citet{Shahbaz2018} and \citet{Papitto2019}. 

We notice that the strong infrared flaring activity is mostly concentrated in the first $\sim$5000s of observation, while the second part of the light curve shows a more quiet behaviour (Fig. \ref{NIR_opt_X_lc}, upper panel). 

\begin{figure*}
\centering
\includegraphics[scale=0.65]{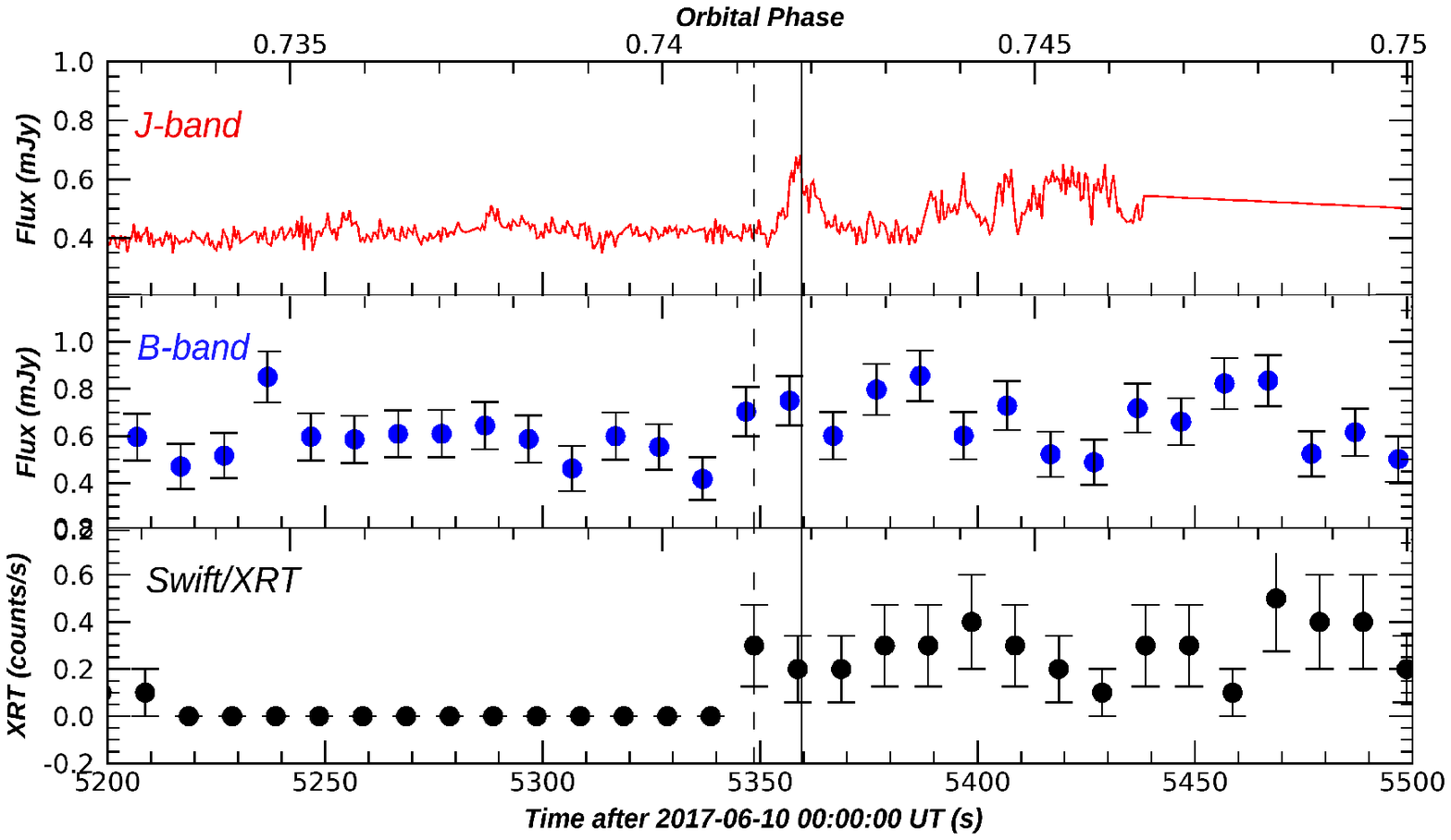}
\caption{From top to bottom, zoom of the NIR, optical and X-ray (\textit{Swift}/XRT) light curves of J1023 around the time of the X-ray low/high transition. With a dashed black line, the exact time of the transition is indicated; with a solid black line, the time corresponding to the centroid of the NIR flare is underlined. The top X axis reports the orbital phases of the system, which have been evaluated starting from the ephemeris of \citet{Archibald2009}, phase 0 being the inferior conjunction of the companion star. }
\label{zoom}

\end{figure*}

The \textit{Swift}/XRT light curve shown in the central panel of Fig. \ref{opt_X_lc} is almost featureless, except for one transition from the \textit{low} to the \textit{high} mode at $\sim$5000 s after 2017-06-10 00:00:00 UT. In Fig. \ref{zoom}, a zoom of the three light curves around the time of the X-ray \textit{low}/\textit{high} transition is shown.
In relation to the \textit{low}/\textit{high} transition, we observe an almost simultaneous increase in the optical flux by a factor of 1.7 in 10s (Fig. \ref{zoom}, middle panel), and the emission of a flare in the NIR. To evaluate the time lag between the two, we considered the time corresponding to the first X-ray point associated to the \textit{high} mode as the time of the \textit{low/high} transition, and the centroid of the Gaussian function which best fits the NIR flare as the time of the emission of the flare. In this way, a lag of $11\pm5$ s was estimated (Fig. \ref{zoom}, top panel), with the infrared delayed with respect to the X-rays.

To have a better understanding of the origin of the $J$-band flare in correspondence to the \textit{low}/\textit{high} transition, we evaluated the slope $\alpha$ of the optical/NIR spectral energy distribution (SED). We defined the slope of the SED starting from the $B$-band and $J$-band fluxes ($F_{B}$ and $F_{J}$, respectively) as follows:

\begin{equation}\label{slope_eq}
\alpha=\frac{\log_{10}\left(F_{B}\right)-\log_{10}\left(F_{J}\right)}{\log_{10}\left(\nu_{B}\right)-\log_{10}\left(\nu_{J}\right)},
\end{equation}

where $\nu_{B}$ and $\nu_{J}$ are the central frequencies of the $B$-band ($6.67\times10^{14}$ Hz) and $J$-band ($2.43\times10^{14}$ Hz) filters. Fluxes have been de-reddened using the coefficients reported in \citet{Baglio2016b}.

\begin{figure*}
\centering
\includegraphics[scale=0.55]{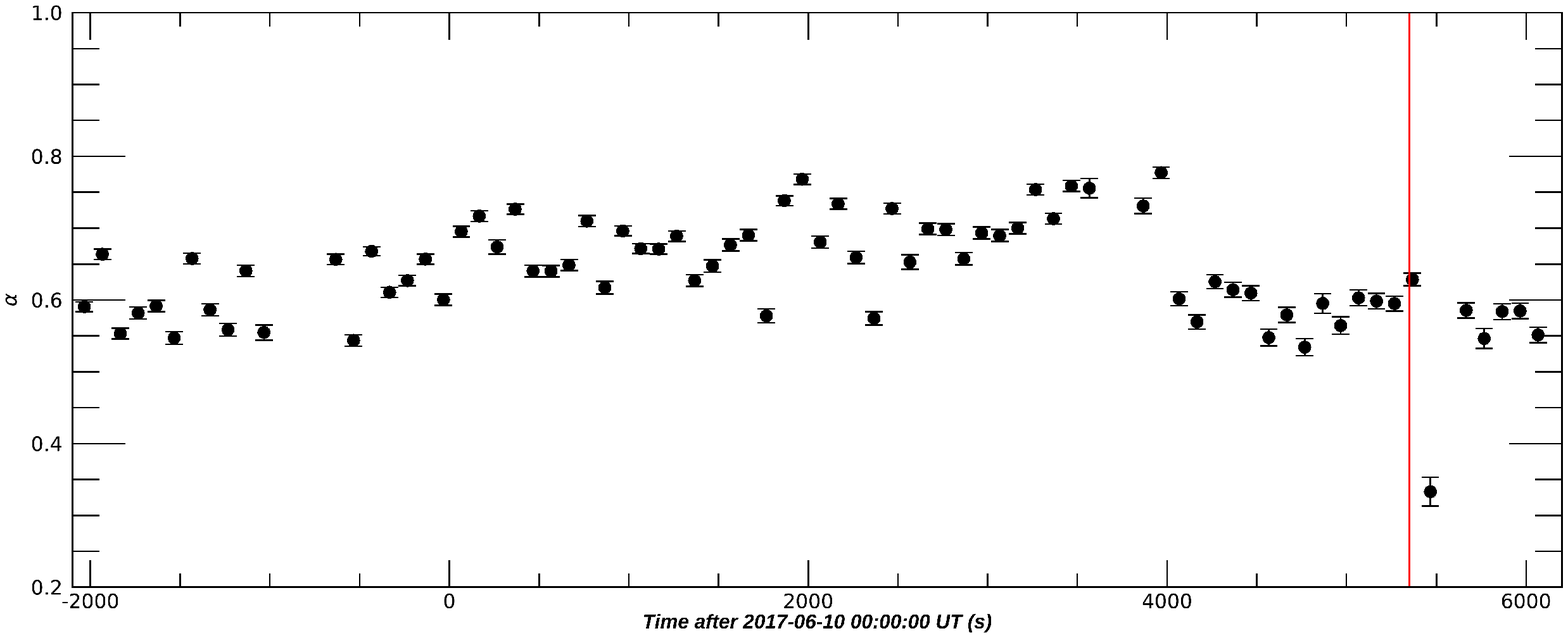}
\caption{Trend with time of the slope $\alpha$ of the optical/NIR spectral energy distribution of PSR J1023, defined as in Eq. \ref{slope_eq}. The solid red line indicates the exact time of the \textit{low}/\textit{high} X-ray transition. The bin time of the curve is 100s. Errors are represented at the 68$\%$ confidence level.}
\label{slope_Fig}

\end{figure*}

The slope of the SED stays approximately constant during the whole duration of our observations (Fig. \ref{slope_Fig}), with a slow increase in the index before 4000s, and a sudden drop after 4000s. No clear variation of the slope in proximity to the infrared flare is however detected.

\subsection{Correlations}
With the aim of revealing any NIR/optical correlation, we tried to evaluate the cross-correlation function (CCF) between the two (NIR and optical) light curves. To do this, only strictly simultaneous optical and NIR data have been considered. The time resolution of the optical light curve is worse than the NIR (10s vs. 0.5s, respectively), so we decided to re-bin the NIR light curve to a 10s time resolution to evaluate the CCF.

The OM light curve shows strong flares at the beginning and at the end of the exposure; the NIR light curve instead has a non-stationary behaviour. As shown in Fig. \ref{NIR_opt_X_lc} (top panel), the first $\sim5000$ s of the NIR light curve are characterized by strong and structured variability, while the last half of the observation does not show signs of strong variability, except for the flare in correspondence to the low/high mode transition at $\sim$ 5000 s after 10/06/2017 00:00:00 UT (Fig. \ref{zoom}, top panel).

\begin{figure}
\centering
\includegraphics[scale=0.35]{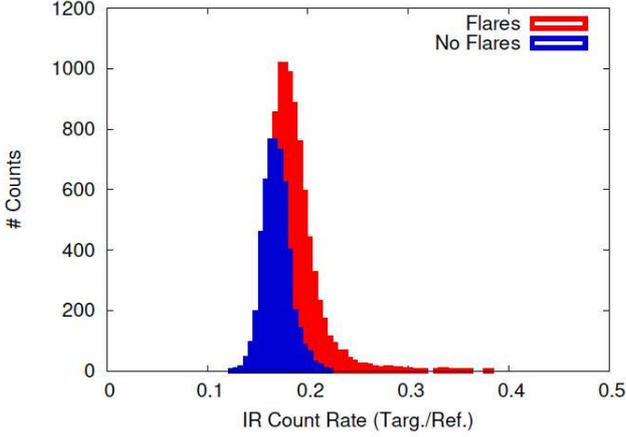}
\caption{NIR light curve distribution of the two regimes. In red, the first, more variable segment is represented, in blue the most stable. Gaussian fits of the two distributions give reduced $\chi^2$ of 1.2 and 4.4 in the quiet and flare part of the light curve with 19 and 29 degrees of freedom, respectively, in accordance with the asymmetry of the flaring part distribution, also visible by eye.}
\label{histo}
\end{figure}

This difference is also evident while computing the distribution of the fluxes for the NIR light curve, i.e. the histogram of the $J$-band count rates values. We computed the histogram of the light curve in these two different regimes: first, we used the first $\sim$5000 s of observation, characterized by strong and structured variability; secondly, we considered the $\sim 1000$ s before the low/high X-ray transition, where the light curve is almost flat, without signs of variability. In the first case, an asymmetric distribution was found, with a long tail towards high fluxes, clearly due to the presence of structured variability. In the second, the distribution is instead symmetric, and can be well fitted with a log-normal distribution (with a reduced $\chi^2$ of 1.02; see eq. 3 in \citealt{Uttley2005}). The two distributions are shown in Fig. \ref{histo}.

\begin{figure}
\centering
\includegraphics[scale=0.55]{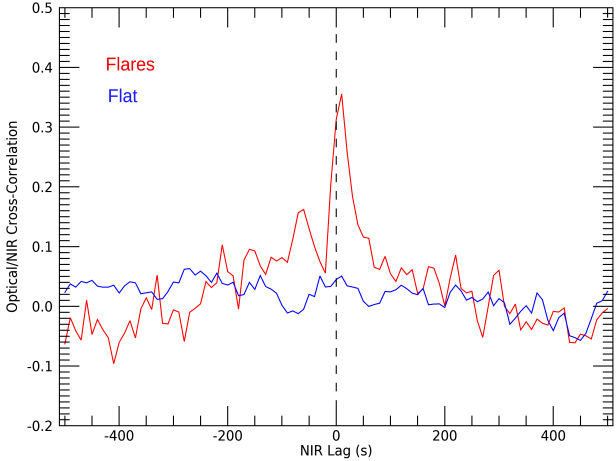}
\caption{Optical/NIR CCF computed in two different selected segments of the NIR light curve. A strong peak is present only in the variable part of the light curve. Uncertainties are of the order of 0.05 ($68\%$ confidence level).}
\label{CCF_Fig}
\end{figure}

Due to this bimodality of the NIR light curve, we decided to compute the optical/NIR CCF using these two separated segments (Fig. \ref{CCF_Fig}). While the second CCF, that is referred to the flat part of the light curve (represented in blue in Fig. \ref{CCF_Fig}), does not show any significant feature, the first CCF (in red in Fig. \ref{CCF_Fig}) presents a clear peak at $7.0\pm 1.3$s. This is an indication that, at least for the part of the light curve which shows flares, the NIR emission has a delay of $\sim 10$s with respect to the optical.

A similar result was also achieved in \citet{Shahbaz2018}, where a narrow, positive peak at $\sim 5$s was detected in the cross-correlation function between the optical $r'$ band and the NIR $K$ band. However, they also detect a broad anticorrelation on $\sim -300$s timescale, and a broad positive correlation at +300s, which we do not observe in our CCF.

\subsection{Power density spectrum}

\begin{figure}
\centering
\includegraphics[scale=0.7]{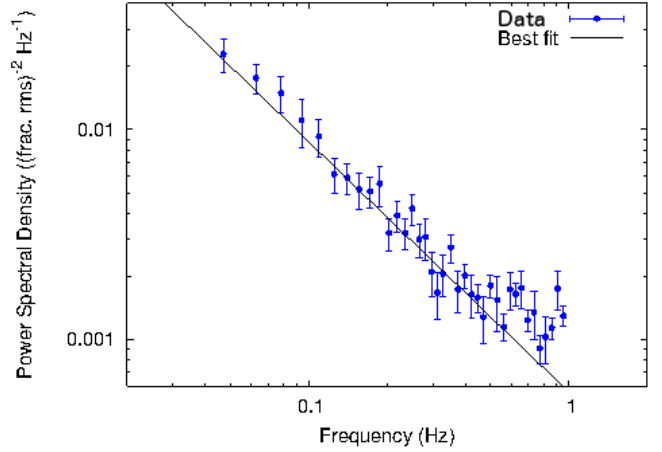}
\caption{The power density spectrum of the $J$-band light curve of PSR J1023+0038. Superimposed, the power-law fit to the PDS is represented (solid line).}
\label{power_spectrum_fig}
\end{figure}

The NIR power density spectrum (PDS) was computed using 128 bin per segment and a logarithmic binning factor of 1.05, allowing to probe frequencies from 0.0625 to 1 Hz. The maximum frequency probed in this work is a factor of 10 higher than the most recent study regarding the IR fast variability of this source \citep{Shahbaz2018}. The PDS shows a decreasing power-law trend with increasing frequencies (Fig. \ref{power_spectrum_fig}), and seems to become flatter for higher frequencies. Following \citet{Shahbaz2018}, we performed a fit between 0.02 and 0.2 Hz with a power-law model. The best fit returned an index of $-1.23 \pm 0.08$, which is consistent with the slope measured by \citet{Shahbaz2018}, and also with the slope measured in the optical by \citet{Papitto2018}, in UV by \citet{Hernandez2016} and in the X-rays by \citet{Tendulkar2014}, within the uncertainties. As shown in Fig. \ref{power_spectrum_fig}, for frequencies higher than 0.6 Hz there is an excess in the PDS. 
However, given the presence of signal near the Nyquist frequency, this excess is spurious and can be explained in terms of aliasing (i.e. the reflection of the power present above the Nyquist frequency also at the frequencies below it; see \citealt{vanderklis1988} for a review). 

This result indicates aperiodic activity and is typically observed in X-ray binaries, both in outburst and quiescence, for which PDS indexes in the range -1 -- -2 have been measured in IR (see \citealt{Shahbaz2018} and references therein).

\section{Discussion}\label{Sec_discussion}

\subsection{On the origin of the $\sim$10s NIR/optical lag}
Studies of the correlations between light curves at different wavelengths are an extremely powerful tool to determine which are the physical processes that are taking place. In particular, cross-correlation studies allow us to determine whether any delay between light curves at different frequencies are present. Fast photometry observations are in this sense crucial if one wants to investigate processes that are occurring on short timescales.
With our campaign, we obtained 0.5s integration time NIR-infrared ($J$-band) observations of PSR J1023+0038, simultaneously with \textit{XMM-Newton}/OM, $Swift$/XRT and XMM-Newton RGS observations. 

Cross-correlation studies of PSR J1023 have already been performed and published. In particular, \citet{Shahbaz2018} show the results of a NIR ($K$-band) and optical ($r'$-band) campaign which revealed a strong, broad anticorrelation at negative lags and a broad positive correlation at positive lags, with a superimposed narrow correlation. They interpret their results as due to reprocessing of the optical emission at infrared frequencies, the narrow positive lag being peaked at $\sim 5$s, which is consistent with the time-scale for the light travel between the binary components, considering a binary separation of $\sim 1.8 \,\rm R_{\odot} $. Moreover, their optical light curve has a resolution good enough to enable them to observe several \textit{active}/\textit{passive} mode transitions, as already reported to occur in the optical by \citet{Shahbaz2015}.

Recently, an X-ray/optical/NIR timing study on the source has been performed \citep{Papitto2019}. In this work, the optical and X-ray correlation has been extensively studied, and the CCFs were evaluated during the X-ray \textit{flaring}, \textit{high} and \textit{low} modes at timescales ranging from 1 to 2.5s over 64 s intervals. X-ray and optical emissions are correlated in all three modes, the optical variability being always delayed with respect to the X-ray variability. In their work, also the connection between NIR and X-ray emission was investigated. The two curves are found to be correlated, with the emission of a slightly lower infrared flux during the X-ray \textit{low} mode with respect to the \textit{high} mode. Moreover, they also found evidence for infrared flares occurring after the transition from the low to the high X-ray mode in three cases, but they do not observe a corresponding flare at the start of any X-ray low mode.

Due to the different time resolution and quality of our datasets, a comparison between our results and those reported in \citet{Shahbaz2018} is difficult to perform. Once the orbital modulation is subtracted (Fig. \ref{opt_X_lc}), we do not observe any transition from the \textit{active} to the \textit{passive} mode (and viceversa) in our $B$-band light curve, in which only erratic variability is detected. Some flares are also observed, but the lack of any simultaneous observation at different wavelength makes an interpretation tricky.
The results reported in Fig. 9 of \citet{Shahbaz2018} might explain the absence of any \textit{active}/\textit{passive} mode transition in our light curve. In fact, according to that analysis, the expected flux difference between the \textit{active} and \textit{passive} mode is supposed to vary depending on the wavelength of the observation, being more pronounced in the $r'$-band (i.e. the band in which \citealt{Shahbaz2018} observed) than in our $B$-band. In particular, we should expect a variation of $\sim 0.1$mJy in the $B$-band, that is comparable with the dimension of the error bars in our light curve (see Fig. \ref{zoom}, mid panel). Therefore, it is almost impossible to detect any \textit{active}/\textit{passive} mode transition in our optical light curve. In addition, the cross-correlation function that we built between our NIR and optical light curves (Fig. \ref{CCF_Fig}) does not show many similarities with the one reported in \citet{Shahbaz2018}. In particular, we observe, for the portion of the light curve where NIR flares are detected (red curve in Fig. \ref{CCF_Fig}), a narrow, positive lag of the NIR with respect to the optical at $7.0\pm1.3$s, which is way higher (almost double) with respect to the one observed in \citet{Shahbaz2018}. Moreover, we do not observe in this CCF any broad correlation or anti-correlation. The blue curve in Fig. \ref{CCF_Fig}, i.e. the one that refers to the part of the light curve where flares are not present, is instead completely featureless, meaning that no correlation can be found between those portions of the optical and NIR light curves.

If we first focus on the flat parts of the light curves (blue curve in Fig. \ref{CCF_Fig}), we can easily convert the values of the fluxes in magnitudes, and calculate a mean value of $15.430\pm0.030$, which is consistent with the mean magnitude in $J$-band reported in the literature for PSR J1023 (see e.g. \citealt{CotiZelati2014} and \citealt{Baglio2016b}). In particular, \citet{Baglio2016b} report the broadband fit of the NIR-X-rays SED, demonstrating that the NIR and optical part of the curve can be well fitted by a combination of the contribution of an irradiated star (which is predominant in the $J$-band) plus the truncated accretion disc, which is the sum of blackbody contributions at different temperatures. We can therefore conclude that, for those portions of the light curve that do not show evidence for flaring activity, what is observed is again a combination of the emission coming from the irradiated companion star and a residual accretion disc. This also explains why no correlation or lag between the NIR and the optical emission can be found in the flat part of the light curves.

A further test for this comes from the independent calculation of the expected $J$-band integrated flux emitted by a G-type companion star, which is irradiated by the spin-down luminosity of an active millisecond radio pulsar and by X-ray luminosity. 

To do this, we consider the estimate of the spin-down luminosity $L_{\rm SD}$ of J1023 given by \citet{Deller2012} and \citet{Archibald2013}, $L_{\rm SD}\sim 4.4\times10^{34} \, \rm erg\, \rm s^{-1}$. 
We can then estimate the size of the system using the equations reported in \citet{Eggleton1983} to compute the Roche lobes dimensions, from which we derive an orbital separation $a$ of $1.3\times 10^{11}\, \rm cm$, i.e. $\sim 1.8\, R_{\odot}$, considering a mass of $0.2\, M_{\odot}$ for the companion star, and $1.7\, M_{\odot}$ for the NS (\citealt{Archibald2009}; \citealt{Deller2012}). We can evaluate the fraction of spin down luminosity which is intercepted by the companion star, under the assumption of isotropic emission of the pulsar and considering the typical geometry of a LMXB. In particular, the fraction of spin down luminosity of the pulsar which is intercepted by the companion star, with a radius equal to its Roche lobe radius $R_{\rm L_{2}}$, will be:
\begin{equation}
 L_{\rm SD, int}= L_{\rm SD}\frac{1-\cos \theta}{2},
 \end{equation}
 where $\theta$ is the angular radius of the companion star as seen from the pulsar, i.e. $\theta=arsin(R_{\rm L_{2}}/a)=0.22 \rm rad$. Therefore, $L_{\rm SD, int}=5.28\times 10^{32} \rm erg/s$, considering an almost negligible albedo.

With a similar reasoning, considering an emitted X-ray luminosity of $7\times 10^{33}\, \rm erg\, s^{-1}$ (obtained by converting the measured counts/sec in our X-ray light curve and considering the known distance of 1.37 kpc; \citealt{Deller2012}), we can also derive the fraction of X-ray luminosity that irradiates the companion, which is $8.4\times 10^{31} \rm erg\, \rm s^{-1}$. The intrinsic luminosity $L_{\rm *}$ of the $G$-type star can instead be evaluated starting from the surface temperature of the companion reported in \citet{Thorstensen2005}, i.e. 5700 K. Using the Stefan-Boltzmann relation, $L_{*}=6.32\times10^{32} \rm erg\, \rm s^{-1}$. 
By summing together these three luminosities, we obtain a maximum luminosity of the star of $1.24\times10^{33}\, \rm erg\, s^{-1}$, which corresponds to a surface temperature of $\sim 6650$K. If we integrate the black-body function for this temperature in the $J$-band, we obtain a black body flux of $\sim 4\times10^{-13}\rm \, erg\,cm^{-2}\,s^{-1}$, which is consistent with the flux that we measure in the flat part of our $J$-band light curve (i.e. $\sim 3\times10^{-13}\rm \, erg\,cm^{-2}\,s^{-1}$). Therefore, we can safely conclude that what we are observing in the flat part of the NIR light curve is the near infrared emission of the irradiated companion star only, which is not expected to lag the optical emission, as observed. We caution however that for this calculation several approximations have been done.
As a consequence of this, one would expect to observe an orbital modulation of the NIR light curve. While a modulation is observed in the $B$-band light curve, this is not apparent in $J$-band. We note, however, that the $B$-band semi-amplitude that we measure is rather low, of the order of $10\%$. This is lower with respect to what was observed in \citet{CotiZelati2014}, where a $g$-band semi-amplitude of the $40\%$ was detected. Assuming a scaling with the wavelength, the $J$-band expected semi-amplitude should be of the order of $5\%$, i.e., comparable to the statistic uncertainty of our NIR dataset, suggesting that a low level modulation is indeed present, although hidden in the uncertainty of our NIR dataset.

Switching to the red CCF in Fig. \ref{CCF_Fig}, i.e. the cross-correlation function which is relative to the portion of the NIR light curve where flaring activity is detected, what we observe is the presence of a narrow peak at 10s lag, which indicates that the NIR is delayed with respect to the optical with a lag of $\sim$10s. To investigate the origin of this positive lag, we first considered the hypothesis of the NIR emission being the result of the reprocessing at the companion star surface of the optical emission coming from the accretion disc, as suggested in \citet{Shahbaz2018}. According to \citet{Bogdanov2015}, the optical variability closely matches the X-ray variability (at least in terms of flares emission, which the authors interpret as phenomena which span from optical to at least X-ray frequencies), and therefore any X-ray flare should in principle give rise to a corresponding flare in the NIR.
Because we do not possess high time-resolution X-ray observations of the target which are contemporaneous to our NIR and optical datasets, we decided to consider the X-ray 0.3-10 keV variations reported in \citet{Bogdanov2015} to prove this hypothesis, under the assumption that the X-ray variations in the light curves will not be too different in different epochs. These variations are on average of $\sim 30\, \rm counts \, \rm s^{-1}$, which, considering a power law model as X-ray spectrum, correspond to an unabsorbed flux of $\sim 4.5\times10^{-11} \, \rm erg\,s^{-1}\, cm^{-2}$. Considering the known distance of PSR J1023 (1.37 kpc), we evaluated a luminosity of $\sim 1\times 10^{34} \, \rm erg\, s^{-1}$, which, considering an orbital separation of $\sim$1.8 $R_{\odot}$, translates into a flux irradiating the companion star of $\sim 5\times 10^{10} \, \rm erg\, cm^{-2}\, s^{-1}$. If we multiply this result for the area of the companion star which is irradiated, we obtain the luminosity which is absorbed by the companion star, i.e. $1.2\times 10^{32} \, \rm erg\,s^{-1}$. Adopting an almost negligible albedo $\eta\sim 0.1$,  and using the Stefan-Boltzmann relation with a surface temperature of the companion star of $\sim 6650$K (that we just evaluated considering the spin-down luminosity and the X-ray luminosity as sources of irradiation), we obtain that such a flux variation in the X-rays would increase the surface temperature of the companion star up to $\sim 7080$K. Integrating the black-body Planck function for these two temperatures over the $J$-band filter wavelengths, we finally obtain that the expected flux variation due to reprocessing of the X-ray/optical flaring emission in this band should be of a factor 1.16. 
However, the amplitude of the flares that we see in our NIR light curve is higher, with a fractional flux variation of a factor of $\sim 1.75$. Therefore, the X-ray band variability of PSR J1023 is not sufficient to explain the NIR flares that we observe. This, together with the clear evidence (see Fig. \ref{NIR_opt_X_lc}) that no optical/X-ray flare is observed during the HAWK-I observations, permits to exclude a reprocessing origin of the NIR light curve (or at least a further component has to be invoked). 
Moreover, the $\sim$10s lag that we observe in our CCF is too high to support a reprocessing origin of the NIR emission, the 1.8$R_{\odot}$ orbital separation being consistent with a $\sim 4\rm s$ lag only (as reported in \citealt{Shahbaz2018}). Therefore, NIR flaring seem to have a different origin with respect to optical/X-ray flares (which are not detected in our light curves, but are observed for example in the work of \citealt{Papitto2019} using OM data) for J1023.

Different scenarios might be considered to explain the lag that we observe between the NIR and optical light curves.
\citet{Bogdanov2018} observe the emission of radio flares from the system, that the authors interpret as due to the emission of collimated jets in correspondence with the $low$ X-ray state. The spectrum of the radio emission is found to be flat or slightly inverted at the lowest frequencies, which is a typical signature of the emission of synchrotron radiation from jets.
Jets from NS X-ray binaries have been observed mostly in X-ray binaries in outburst or in persistent systems (\citealt{Migliari2006}; \citealt{Russell2007}; \citealt{Russell08}; \citealt{Migliari10}; \citealt{Baglio2016a}; \citealt{Tudor2017}). Therefore, we do not possess an extensive literature in case of jets emitted in quiescence, neither in the particular intermediate state in which J1023 lingers.
Several cross-correlation studies between X-ray, optical and NIR light curves have been performed for black hole (BH) X-ray binaries (see for example the results for GX 339-4 and V404 Cyg: \citealt{Casella2010}; \citealt{Kalamkar2016}; \citealt{Gandhi2017}; \citealt{Vincentelli2018}).
For these systems, the NIR has always been found to be delayed with respect to the X-ray emission, with a typical lag of $\sim 0.1\rm s$. Such lag has been explained in terms of the presence of a strongly variable accretion flow, that emits in the X-rays, which is able to inject velocity fluctuations at the base of the jet, driving internal shocks at large distances from the BH (\citealt{Malzac2013}; \citealt{Malzac2014}; \citealt{Drappeau2017}; \citealt{Malzac2018}). At the shocks, leptons are accelerated, giving rise to variable synchrotron emission, which should be observed in the radio/NIR. However, according to this model delays between X-rays and NIR of $\sim 0.1$s are expected, as also supported by the observations. Being the delay measured for PSR J1023 $\sim 100$ times higher, this scenario seems to be excluded.

\begin{figure}
\centering
\includegraphics[scale=0.7]{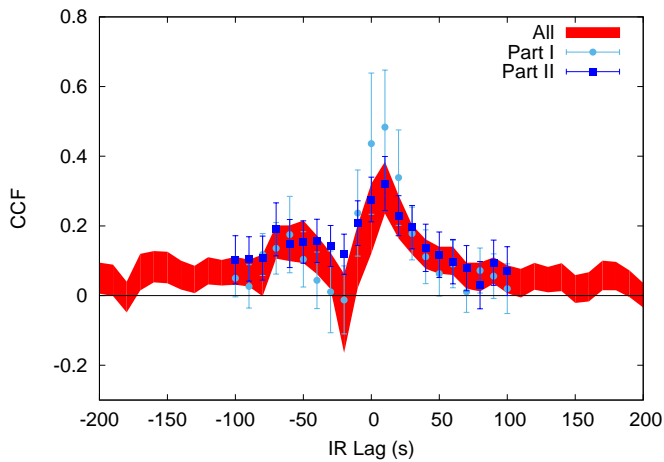}
\caption{Optical/NIR CCF obtained in the following two segments: the first $\sim 2000$s of observation, and the following $\sim 2500$s of observation (light blue dots and blue squares, respectively). Superimposed, the CCF built considering both segments together. No significant differences can be reported.}
\label{CCF_comparison}

\end{figure}

Another intriguing possibility is instead that the NIR emission might be delayed with respect to the optical due to reprocessing of the optical emission in a region that is found outside the system itself (if the optical signal propagates at the speed of light, a 10s delay implies a distance of $\sim3\times 10^{11}$cm from the region of the optical emission, i.e. $\sim 4 R_{\odot}$, which has to be outside the system)\footnote{We caution that in principle, without a knowledge of the angle subtended at the source by the earth and the reprocessing site, it is impossible to put a firm constraint on the distance between the source and the reprocessor, although, as an order of magnitude, we can give some approximate estimates. Therefore the reader should consider that the numbers presented in this discussion contain this uncertainty.}. 
The recent observations reported in \citet{Papitto2019} seem to suggest that a rotation-powered pulsar may always be active in the system, even if the source is in its accretion state. In the \textit{high} mode, the relativistic wind emitted from the radio pulsar, interacting close to the light cylinder radius with the flow of matter which accretes from the companion star, creates a shock where electrons are accelerated, therefore producing synchrotron radiation in the form of optical and X-ray pulsations, which are observed for J1023. 
If the radio pulsar is on, the momentum that the pulsar radiation exerts onto the matter that reaches the inner Lagrangian point of the system causes an inhibition of the accretion process. This might create a stream of matter which, instead of being accreted onto the compact object, is ejected, and forms a spiral around the entire system. This phase of ejection of matter is called ``radio-ejection'' phase, and has already been postulated to explain the behaviour of other millisecond pulsar binaries, as in the case of PSR J1740-5340 (\citealt{Burderi2001}; \citealt{Burderi2002}). 
If we consider the optical emission coming from the light cylinder radius due to the interaction between the pulsar wind and the internal part of the accretion disc (as suggested in \citealt{Papitto2019}), the NIR emission might be the reprocessing of the optical emission after the optical photons meet the stream of ejected matter.
If this is true, the 10s delay would give an idea of the projected distance between the region of the optical emission, that is located at some $10^2$ km from the pulsar, and the stream of matter, i.e. $\sim 4 R_{\odot}$.
This hypothesis is also supported by the fact that we do not observe the 10s delay during the whole light curve, which suggests for some kind of asymmetric distribution of the flow of matter around the system. In particular, the 10s lag is observed between phase $\sim 0.3$ and $\sim 0.6$, i.e. in the first $\sim 5000$s of the NIR light curve (see Fig. \ref{NIR_opt_X_lc}). As a test, we tried to compute new CCFs by dividing into two segments the flaring part of the NIR light curve (first segment: the first $\sim 2000$s observation in the $J$-band light curve; second segment: the following $\sim 2500$s of observation), but no significant differences were found, the peak of the CCFs being always at $\sim 10$s (see Fig. \ref{CCF_comparison}). 

\subsection{NIR flares exiting the \textit{low} mode: evidence of jets/outflows?}

\citet{Bogdanov2018} reported on the detection of radio flares emitted in correspondence with the whole duration of an X-ray \textit{low} mode. During a typical \textit{low} mode, the radio spectrum of the flare is observed to evolve from being inverted ($\alpha \sim 0.4$, where $F_{\nu} \propto \nu^{\alpha}$) to steep ($\alpha \sim -0.5$) over several minutes. This enhanced radio flux, characterized by an evolving radio spectrum, is interpreted as a signature of the launching of an evolving outflow/jet.
During the \textit{high} mode, the radio flux drops by a factor of 3-4 with respect to its level during the \textit{low} mode. The radio spectrum in this case is found to be on average slightly inverted ($\alpha \sim 0.2$). 
The fact that the radio emission is not switched off during the \textit{high} mode suggests that the outflow is not totally quenched after the transition to the \textit{high} mode, but a self-absorbed, compact jet might still be emitted.

\begin{figure}
\centering
\includegraphics[scale=0.55]{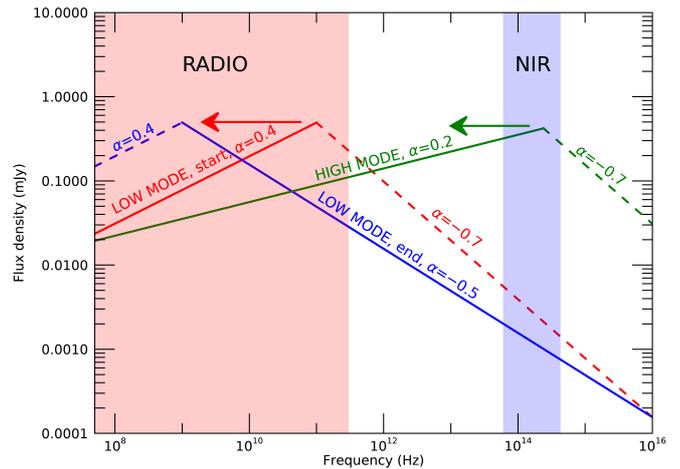}
\caption{Representation of the possible evolution of the outflow emission spectrum during the \textit{high} and the \textit{low} X-ray mode. In red, the possible synchrotron spectrum of the outflow at the start of the \textit{low} mode is represented. In blue, we represented the possible spectrum of the outflow at the end of the \textit{low} mode. In green, the possible spectrum of the outflow at the beginning of the \textit{high} mode is represented. 
All the break frequency values have been chosen arbitrarily, with the only scope of explaining the model, but the spectral index values and the normalization (flux) values are based on observations. With two arrows, the plausible direction of the evolution of the break frequency is sketched. The red and blue regions of the figure represent the radio and NIR bands, respectively.}
\label{jets_fig}

\end{figure}

Jets emitted in low luminosities X-ray binaries, like PSR J1023, are not expected to have a significant contribution in the NIR; in fact, as shown in \citet{Baglio2016b}, no contribution from synchrotron emission of a jet or a generic outflow is observed at NIR frequencies in the broadband SED.
If synchrotron radiation is emitted from an outflow during the X-ray low mode, we can therefore hypothesize that the break frequency of its spectrum will be within the radio band, which means that the jet will mainly contribute in the radio. The fact that the radio spectrum of the jet during the low mode is found to evolve from inverted to steep suggests that the jet break frequency might be found already in the radio band (but at a frequency $>12\rm GHz$) at the beginning of the flare, and then moves to lower frequencies ($<8\rm GHz$) while the jet evolves during the \textit{low} mode. To visualize the evolution of the spectrum, we represented in Fig. \ref{jets_fig} in red the possible synchrotron spectrum of the outflow at the start of the \textit{low} mode, using the slope $\alpha=0.4$ measured by \citet{Bogdanov2018} for the self-absorbed part (solid line), and an arbitrary slope $\alpha=-0.7$ for the optically thin part of the spectrum (dashed line), which is typical of jets and outflows in X-ray binaries. In blue, we represented the possible spectrum of the outflow at the end of the \textit{low} mode, using the slope $\alpha=-0.5$ measured by \citet{Bogdanov2018} for the optically thin part (solid line) and maintaining the slope $\alpha=0.4$ for the optically thick part (dashed line).
After the transition to the \textit{high} mode, a compact self absorbed jet might be launched \citep{Bogdanov2018}. In Fig. \ref{jets_fig}, the possible spectrum of the outflow in the \textit{high} mode is represented in green, using the slope $\alpha=0.2$ reported in \citet{Bogdanov2018} for the self-absorbed synchrotron part (solid line) and the typical $\alpha=-0.7$ slope for the optically thin one (dashed line).
As shown in this representation, if at the beginning of the \textit{high} mode (i.e. when the compact jet is launched) the jet break frequency is in the NIR range, this might give rise to an enhanced NIR flux. We can then suppose that the break frequency shifts towards lower frequencies (but remaining $>12\rm GHz$), without changing the radio spectral shape, in accordance with \citet{Bogdanov2018} (see Fig. \ref{jets_fig}). In this way, we would observe a NIR flare corresponding to the transition to the high mode.
A similar phenomenon was also observed in \citet{Papitto2019}, where an infrared flare was possibly detected in correspondence to three \textit{low-high} transition. In addition, \citet{Papitto2019} did not observe NIR flares when the source enters the low-mode; this is compatible with our interpretation of the NIR flare at the beginning of the X-ray \textit{high} mode as due to the evolving position of the jet-break frequency from higher (NIR) to lower (radio) frequencies during the duration of the \textit{high} mode. 

\section{Conclusions}\label{Sec_conclusions}
In this paper we reported on a near infrared (VLT/HAWK-I, $J$-band, 0.5 s time resolution), optical (\textit{XMM-Newton}/OM, $B$-band) and X-ray ($Swift$/XRT and \textit{XMM-Newton} RGS) simultaneous campaign performed on 9-10 June 2017 on the transitional millisecond pulsar PSR J1023+0038. The main results of this campaign are summarized here:

\begin{itemize}
\item The optical light curve shows the expected sinusoidal modulation at the source 4.75 hr orbital period; once the sinusoidal modulation is subtracted, no evidence for flaring activity or transitions between the $active$ and $passive$ mode is detected. The NIR light curve instead is found to be extremely variable, with the presence of strong flares of $\sim 0.3$mJy amplitude and duration of $\sim minutes$ in the first $\sim 5000$s of observation, while the rest of the light curve is found to be mostly flat, except for one single strong flare. The X-ray light curve shows one transition from the $low$ to the $high$ X-ray mode, which is found to happen soon before one of the strongest NIR flares that we detect in the NIR light curve.

\item We computed the cross-correlation function between the optical and NIR light curves. We divided the NIR and optical light curves in two different segments, due to the bimodality of the NIR light curve, and computed two different cross-correlation functions. The CCF which was relative to the flat part of the NIR light curve is flat, and no significant feature is reported. The one relative to the flaring segment of the curve instead shows a clear peak at $\sim$10s, which indicates that the NIR is delayed with respect to the optical with a 10s lag. The power density spectrum shows a decreasing power-law trend with index $\sim -1.2$, which typically indicates the presence of some aperiodic activity in X-ray binaries.

\item We tentatively interpret the 10s lag between optical and NIR as due to reprocessing of the optical emission at the light cylinder radius with a stream of matter which spirals around the system after being ejected due to the radio-ejection phase. This interpretation is strongly supported by the high time delay, which is a clear indication that the reprocessing of the optical emission cannot happen in a region which is found inside the system. 

\item We observed a NIR flare in correspondence to the transition from the $low$ to the $high$ X-ray mode of the source. The interpretation of this feature is non-trivial. We hypothesize that the flare might be due to the emission of a jet or a non-collimated outflow, as was already suggested by \citet{Bogdanov2018}. According to our interpretation, the infrared flare would be due to the evolving spectrum of the jet/outflow, the jet-break frequency moving rapidly from higher (NIR) to lower (radio) frequency after the launching of the jet, which has to occur at the $low/high$ mode transition.

\end{itemize}

\begin{acknowledgements}
The research reported in this paper is based on observations made with ESO Telescopes at the Paranal Observatory under programme ID 299.D-5023 (PI: Campana).
PDA and SC acknowledge support from ASI grant I/004/11/3.
\end{acknowledgements}


\end{document}